\documentstyle[12pt]{article}
\begin{document}
\baselineskip 24pt
\begin{flushright} \ February \ 1998\\
SNUTP 98-005 \\
\end{flushright}
\begin{center}
{\large \bf Integrable Extension of Nonlinear Sigma Model }
\vspace{.5cm}

Phillial Oh $^{1}$    \\
 {\it Department of Physics, Sung Kyun Kwan University,
Suwon 440-746,  Korea }\\

\end{center}

\vspace{1.0cm}

\begin{center}
{\bf Abstract} \\
\end{center}
We propose an integrable extension of nonlinear sigma model  
on the target space of Hermitian symmetric space (HSS). 
Starting from a discussion of soliton solutions of $O(3)$ model and an 
integrally extended version of it, we construct general theory defined
on arbitrary HSS by using the coadjoint orbit method.  
It is based on the exploitation of a covariantized canonical 
structure on HSS. This term results in an additional first-order
derivative term in the equation of motion, which accommodates the zero 
curvature representation. Infinite  conservation laws of nonlocal  charges 
in this model are derived.
\\

\noindent PACS codes: 11.10.Lm, 11.10.Kk\\

\noindent
\hbox to 10cm{\hrulefill}
\baselineskip 12pt

\noindent $^1$ {\small
E-mail: ploh@newton.skku.ac.kr}

\thispagestyle{empty}
\pagebreak

\baselineskip 24pt


The study of integrability of the nonlinear sigma model (NLSM) on the
target space of homogeneous space $G/H$  was initiated in the work of  
\cite{pohl} where it  was discovered that $O(N)$ NLSM admits a zero
curvature representation, and infinite family of local and nonlocal 
conserved charges \cite{lusc} was derived along with one parameter family 
of B\"acklund transformation. 
This result was immediately  generalized to the target space of
$CP(N)$ \cite{eich} and to the principal chiral $SU(N)$ model where
nontrivial soliton solutions were obtained by applying the inverse scattering
technique \cite{zakh}. Later, the complete integrability on the target space of
Riemannian symmetric spaces was established through the works of
\cite{eich1}, and various aspects of integrability were developed and applied to 
the survey of two dimensional quantum field theory \cite{abda}.   

In this Letter, we search for a possible integrable extension of the
NLSM on the homogeneous space $G/H$. Our main result is that the NLSM allows
an integrable extension when the target space is given by the Hermitian
symmetric space (HSS) which is a symmetric space equipped with a complex structure
acting on the coset \cite{helg}. In achieving this, a covariantized canonical
structure term on HSS is added to the original action which 
results in a first-order total derivative in the 
equation of motion, and this is found to be completely integrable. 
In the process, we use the coadjoint orbit method \cite{oh}
as an essential tool. 
We start from a discussion of an integrable extension of 
the simple $O(3)$ model in this method
and its soliton solution to give some motivation. 

Let us consider the action of $O(3)$ NLSM: 
\begin{equation}
S=\frac{1}{2}\int d^2x\partial_\mu\vec{Q}\cdot
\partial^\mu\vec{Q}
,~~ \vec{Q}\cdot\vec{Q}=1.\label{action}
\end{equation}
We consider solution of the following equation;
(with  prime $= \frac{\partial}{\partial x},$~
dot $= \frac{\partial}{\partial t})$
\begin{equation}\label{eqmotion}
\vec Q\times
\vec Q^{\prime\prime}
-\vec Q\times
\ddot{\vec Q}=0.
\end{equation}
Note that cross product with $\vec Q$ reproduces the formal equation 
of motion of (\ref{action}), 
$\Box \vec Q-(\vec Q\cdot \Box \vec Q)\vec Q=0.$
We use the standard parameterization for $S^2$ given by 
$\vec{Q}=(\sin{\theta}\cos{\phi},\sin{\theta}\sin{\phi},
\cos{\theta})$.
Substitution into (\ref{eqmotion}) leads to
\begin{eqnarray}
2\cos\theta\theta^\prime\phi^\prime+\sin{\theta}\phi''
-2\cos\theta\dot{\theta}\dot{\phi}
-\sin\theta\ddot\phi&=&0,\\
\theta''-\sin\theta\cos\theta\phi^{'2}-
\ddot\theta+\sin\theta\cos\theta\dot\phi^2&=&0.
\label{aaaa}
\end{eqnarray}

Let us try solutions of the form 
$\theta\equiv\theta{(x-vt)},
~\phi\equiv\hat{\phi}(x-vt)+\Omega t$.
We obtain 
\begin{eqnarray}
0 &=& 2(1-v^2)\cos\theta\theta_\xi\hat\phi_\xi
+(1-v^2)\sin\theta\hat\phi_{\xi\xi}
+2\Omega v\cos\theta\theta_\xi, \label{aaaaa}\\
0 &=& (1-v^2)\theta_{\xi\xi}
-(1-v^2)\sin\theta\cos\theta\hat\phi_\xi^2
-2v\Omega\sin\theta\cos\theta\hat\phi_\xi
+\Omega^2\sin\theta\cos\theta, \label{bbbb}
\end{eqnarray}
with $\xi=x-vt$.  
Let $\chi=(1-v^2)\sin\theta\hat\phi_\xi~(\vert v\vert<1)$.
Then, from (\ref{aaaaa})
\begin{equation}
\chi_\xi = -\theta_\xi (2v\Omega \cos\theta +
\chi \cot \theta ),
\end{equation}
which upon integration yields
\begin{equation}
\chi = \frac{v\Omega (\cos 2\theta + c_0) }{2\sin\theta}.
\end{equation}
From (\ref{bbbb}), we obtain
\begin{equation}
{\theta_\xi}^2 + \frac{\chi^2 }{(1-v^2 )^2} 
- \frac{\Omega^2\cos2\theta }{2(1-v^2)}= E.
\end{equation}
We choose $c_0 = -1$, which simplifies  $\chi$ to 
$\chi=-v\Omega \sin \theta$.

Let us consider the effective potential ${\theta_\xi}^2 + 
V_{eff}(\theta)= E$
given by 
\begin{equation}
V_{eff}(\theta)  
=- \frac{\Omega^2\cos^2\theta}{(1-v^2)^2} 
+ \frac{\Omega^2(1+v^2)}{2(1-v^2)}.
\end{equation}
For a soliton solution, we choose $E=\frac{\Omega^2(1+v^2)}{2(1-v^2)}$,
which yields
\begin{equation}
{\theta_\xi}^2
=\frac{\Omega^2}{(1-v^2)^2}\cos^2\theta.
\label{sine}
\end{equation}
Note that differentiation 
of the above equation 
with respect to $\xi$ leads to
the sine-Gordon equation in terms of $\bar\theta\equiv 2\theta$
\cite{pohl}.
One can show that the solution  is given by
\begin{equation}
\sin\theta=\pm\tanh\beta(\xi-\xi_0),
\label{cru}
\end{equation}
with $\beta=\frac{\Omega}{1-v^2}$.
Hence we obtain
\begin{eqnarray}
\theta(x,t) &=& \sin^{-1}\left[\pm
\tanh\left(\frac{\Omega}{1-v^2}(x-vt-\xi_0)\right)\right],
\label{wish}\\
\phi(x,t) &=& \phi_0-\frac{v\Omega}{1-v^2}(x-vt)+\Omega t.
\end{eqnarray}
We find that $\Omega=0$ produces only trivial solution, and $v=0$
gives the stationary soliton solution. The soliton is of topological
nature, because $\sin\theta\rightarrow \mp 1,$ as 
$x\rightarrow\mp\infty$ or $x\rightarrow\pm\infty$ \cite{com}. 

Motivated by the above procedure,
we search for a simple possible integrable extension in the form
\begin{equation}\label{extension}
\vec{Q}\times{\ddot{\vec Q}}
-\vec{Q}\times
\vec{Q}^{\prime\prime}
=\gamma^0{\dot{\vec Q}}+\gamma^1\vec{Q}^\prime,
\label{eqt}
\end{equation}
where $\gamma^0$ and $\gamma^1$ are constants. This form is suggested 
by an observation that with $\gamma^1=0$ and in absence of second order
time derivative term in the left hand side, the above is the 
continuous Heisenberg ferromagnet equation which is well-known to be
completely integrable \cite{fadd}. In terms of $\xi=x-vt$, the above
equation has a good chance of being integrable. 

Again let us use 
$\vec{Q}=(\sin\theta\cos\phi,\sin\theta\sin\phi,\cos\theta)$
and try solution of the form $\theta\equiv\theta(\xi),
~\phi\equiv\hat\phi(\xi)+\Omega t$. Substitution into (\ref{eqt}) gives
\begin{eqnarray}
-v\gamma^0\theta_\xi+\gamma^1\theta_\xi
&=&2(1-v^2)\cos\theta
\theta_\xi\hat\phi_\xi+(1-v^2)\sin\theta\hat\phi_{\xi\xi}
+2v\Omega \cos\theta\theta_\xi,
\label{aequation}\\
v\gamma^0\sin\theta\hat\phi_\xi&-&\gamma^0\Omega\sin\theta
-\gamma^1\sin\theta\hat\phi_\xi
=\Omega^2\sin\theta\cos\theta\nonumber\\
&+&(1-v^2)\theta_{\xi\xi}
-(1-v^2)\sin\theta\cos\theta{{\hat\phi}_\xi}^2
-2v\Omega\sin\theta\cos\theta\hat\phi_\xi.
\label{bequation}
\end{eqnarray}
Let $\chi=(1-v^2)\sin\theta\hat\phi_\xi~(\vert v\vert<1)$ as before. 
Then, from (\ref{aequation}) we find
\begin{equation}
\chi=\frac{v\Omega \cos 2\theta-2\Gamma\cos\theta+c}
{2\sin\theta},
\end{equation}
with $\Gamma=\gamma^1-v\gamma^0$.
(\ref{bequation}) gives
\begin{equation}
E={\theta_\xi}^2
+\frac{\chi^2}{(1-v^2)^2}
-\frac{\Omega^2\cos 2\theta}{2(1-v^2)}-
\frac{2\Omega\gamma^0\cos\theta}{(1-v^2)}.
\label{effect}
\end{equation}

With  $\Gamma=2\Omega v,~c=3\Omega v$, we have
\begin{equation}
\chi=\frac{\Omega v {(\cos\theta-1)}^2}{\sin\theta}.
\end{equation}
And the effective potential is given by (\ref{effect}) as
\begin{equation}
V_{eff}(\theta)=
\frac{\Omega^2v^2}{{(1-v^2)}^2}
\frac{{(1-\cos\theta)}^2}{1+\cos\theta}
[(1-\cos\theta)-\frac{1-v^2}{v^2}(1+\cos\theta)]+
\frac{3\Omega^2}{2(1-v^2)},
\end{equation}
where we have set $\gamma^0=-\Omega$.
It can be put into the following expression;
\begin{equation}
V(\theta)=\frac{4\Omega^2}{{(1-v^2)}^2}
\frac{\sin^4\frac{\theta}{2}}{\cos^2\frac{\theta}{2}}
[\sin^2\frac{\theta}{2}-(1-v^2)]+\frac{3\Omega^2}{2(1-v^2)}.
\end{equation}
Defining $\beta\equiv\frac{\Omega}{1-v^2},
\alpha^2\equiv (1-v^2)$ and choosing $E=\frac{3\Omega^2}{2(1-v^2)}$, 
we find
\begin{equation}
{\theta_\xi}^2+4\beta^2
\frac{\sin^4\frac{\theta}{2}}{\cos^2\frac{\theta}{2}}
(\sin^2\frac{\theta}{2}-\alpha^2)=0.
\label{equate}
\end{equation}
Let $\omega=\sin\frac{\theta}{2}$. Then, (\ref{equate}) becomes
\begin{equation}
 d\omega=\mp\beta \omega^2\sqrt{\alpha^2-\omega^2}d\xi,
\end{equation}
which can be immediately integrated to yield
\begin{equation}
\omega=\pm\frac{\alpha}{\sqrt{1+\alpha^4\beta^2(\xi-\xi_0)^2}}.
\end{equation}
Hence the soliton solution is given by
\begin{equation}
\theta(x,t)=2\sin^{-1}\left[\pm\sqrt{\frac{1-v^2}
{1+\Omega^2{(x-vt-\xi_0)}^2}}\right].
\end{equation}
Also from $\chi=(1-v^2)\sin\theta\hat\phi_\xi
=\Omega v\frac{{(1-\cos\theta)}^2}{\sin\theta}$, we have
\begin{equation}
\phi(x,t)=\phi_0+\tan^{-1}\left[\left(\frac{\Omega}{v}
\right)(x-vt-\xi_0)\right]+\Omega t.
\end{equation}
Unlike the previous case, this soliton is non-topological:
$\sin(\theta/2)$ starts at $0$ at $\xi=-\infty$, reaches
$\pm\sqrt{1-v^2}$ at $\xi=\xi_0$, and goes back to $0$ at $\xi=\infty$. 

The above analysis suggests that the  first order derivative term in
the right hand side of (\ref{eqt}) could be one possible integrable
extension of $O(3)$ model. It originates from the  covariantized 
symplectic structure on $S^2$ and can be generalized to arbitrary HSS, 
as can be seen from the canonical structure of integrable generalized 
Heisenberg ferromagnet system on HSS \cite{oh}. 
To describe a general formalism on HSS, we start with a brief summary of
the coadjoint orbit approach to NLSM \cite{oh}. 
Consider a group $G$, Lie algebra ${\cal G}$
and its dual ${\cal G}^*: X \in {\cal G};~u \in 
{\cal G}^*$. Let us assume that the inner product is given by the 
trace: $<u,X>=\mbox{Tr}(Xu)$. Then, ${\cal G}$
and ${\cal G}^*$ are isomorphic and the coadjoint
orbit,  which is generated as the orbit of coadjoint action 
of the group $G$ can be parametrized by
\begin{equation}
Q=gKg^{-1}=Q^At^A;~K\in{\cal G}~ 
(A=1,\cdots, \mbox{dim}{\cal G}),
\end{equation}
where $t^A$'s are the generators of  ${\cal G}$.
One can see that $Q$ on each coadjoint orbit characterized by
the element $K$ is defined on a homogeneous space $G/H$, 
where $H$ is the stabilizer of the point 
of $K$. It is well-known that
there is a natural symplectic structure on each orbit,
which comes from the cotangent bundle $T^*G\cong G\times {\cal G}^*$ 
via symplectic reduction \cite{abra}. The canonical one-form on $G/H$ 
which has a  left global $G$ and a right local 
$H$ invariance is given by  
\begin{equation}
\Theta=<\mbox{Ad}^*(g)K,\delta gg^{-1}>=
\mbox{Tr} (Kg^{-1}\delta g).
\label{nrel}
\end{equation}

Based on the above, we propose 
the following action for the NLSM on the target space of 
coadjoint orbit:
\begin{equation}
S=\int d^2x \mbox{Tr}\left[\partial_\mu Q \partial^\mu Q
+ 2 \gamma^\mu(K g^{-1}\partial_\mu g)
 \right],\label{intac}
\end{equation}
where $\gamma^\mu$ is a two-vector.
Note that the above action has a right local  $H$ symmetry 
which is responsible for the reduction to the coset space $G/H$.
The first term becomes the standard Lagrangian of NLSM on $G/H$
\cite{oh}; 
$\sim g_{\alpha\beta}\partial_\mu\bar\psi^\alpha  
\partial^\mu\psi^\beta$, where $g_{\alpha\beta}$ is the metric
in terms of the local coordinate $\psi^\alpha$ on $G/H$.
The second term is the covariantized canonical structure of (\ref{nrel})
on the coadjoint orbit. In a frame with $\gamma^0=1$ and $\gamma^1=0$, 
it  becomes the canonical structure $\sim p
\dot q$. The  equation of motion with respect to the variation
of $g$ is given by
\begin{equation}
\partial_\mu [Q,\partial^\mu Q]+\gamma^\mu \partial_\mu Q=0.
\label{eqmt}
\end{equation}
Note that in $SU(2)$ case with $Q=\vec Q\cdot \vec t$, 
the above equation 
reduces to (\ref{extension}).

To demonstrate a complete integrability
of (\ref{eqmt}) on HSS \cite{helg}, we first explain HSS in terms of
coadjoint orbit language. 
Let us recall that  symmetric space is a coset space 
$G/H$ for Lie groups 
whose associated Lie algebras ${\cal G}$ and $ {\cal H}$, 
with the decomposition
${\cal G} = {\cal H} \oplus {\cal M}$, satisfy the commutation
relations,
\begin{equation}
[{\cal H} , ~ {\cal H}] \subset {\cal H}, ~~ [{\cal H}, ~ {\cal M}]
\subset
{\cal M}, ~~ [{\cal M},~ {\cal M}] \subset {\cal H} .
\label{algebra}
\end{equation}
For HSS, (i) the element $K$ is chosen as the central element of 
the Cartan subalgebra of ${\cal G}$
whose centralizer in ${\cal G}$ is $H$. 
(ii) Then,  we have
$J = \mbox{Ad} (K)$ acting on the coset 
is a linear map satisfying the complex
structure condition $J^{2} = -1$, which along with
(\ref{algebra}) gives the identity
\cite{oh}:
\begin{equation}
[Q,[Q,\partial_\mu Q]]=-\partial_\mu Q.
\label{use}
\end{equation}

Next, we give the zero curvature representation of (\ref{eqmt}).
Introduction of
\begin{equation}
\label{letusintroduce}
A_\mu =a[Q,\partial_\mu Q]+b\epsilon_\mu^{~\rho}
[Q,\partial_\rho Q]+c_\mu Q,
\label{one}
\end{equation}
and substitution into
\begin{equation} 
\epsilon^{\mu\nu}\left(\partial_\mu A_\nu -
\partial_\nu A_\mu+[A_\mu,A_\nu ]\right)=0,
\end{equation}
with use of (\ref{use}) for HSS yield
\begin{eqnarray}
0&=&2b\partial_\mu [Q,\partial^\mu Q]
+2\epsilon^{\mu\nu}c_\nu\partial_\mu Q
+2a\epsilon^{\mu\nu}c_\nu\partial_\mu Q
-2bc^\mu\partial_\mu Q\nonumber\\
&&+(a^2+2a-b^2)\epsilon^{\mu\nu}
[\partial_\mu Q,\partial_\nu Q].
\end{eqnarray}
We have ( with $\epsilon^{01}=1,
~\eta_{\mu\nu}=(-,+),
~\epsilon^{\mu\nu}{\epsilon_\nu}^\sigma =\eta^{\mu\sigma})$
\begin{equation}
a^2+2a-b^2=0,~
\epsilon^{\mu\nu}(1+a)c_\nu -bc^\mu =b\gamma^\mu.
\label{wehave}
\end{equation}
Writing ${(a+1)}^2-b^2=1$, we have
\begin{equation}
a=\frac{2}{\lambda^2-1},~~~
~b=\frac{2\lambda}{\lambda^2-1}.
\label{two}
\end{equation}
From the second equation of (\ref{wehave}), we find
\begin{equation}
c_\mu=\frac{4\lambda^2}{(\lambda^2-1)^2}\gamma_\mu+
\frac{2\lambda(\lambda^2+1)}{(\lambda^2-1)^2}
\epsilon_\mu^{~\nu}\gamma_\nu.
\label{three}
\end{equation}
Note that in the case of $c^\mu=\gamma^\mu=0$, the condition
(\ref{wehave}) is precisely the one emerging from the zero
curvature condition of NLSM on the  homogeneous symmetric target 
space \cite{eich1}. 

Permitting the zero curvature condition, (\ref{eqmt}) has
an infinite number of local \cite{fadd} and nonlocal \cite{lusc}
conservation laws.
We compute the nonlocal conserved charge here by making  use of
the well-known technique from the analysis of the principal
chiral model \cite{deve,brez,ogie}.
Given the linear problem,
\begin{equation}
(\partial_\mu+ A_\mu)\psi(x;\lambda)=0,
\label{four}
\end{equation}
let us consider the Laurent expansion around $\frac{1}{\lambda}$,
\begin{equation}
\psi(x;\lambda)=\sum_{n=0}^{\infty}\frac{\psi_n}{\lambda^n},
~~~~\psi_0=1.
\end{equation}
The infinitely many conserved currents are defined as
\begin{equation}
J_\mu^{(n)}=\epsilon_{\mu\nu} \partial^\nu \psi_n(x),
\end{equation}
which automatically satisfies the current conservation
$\partial_\mu J^\mu=0$. The $\psi_n$'s are determined recursively
as follows: Substituting (\ref{one}), (\ref{two}), and (\ref{three})
into
(\ref{four}), one obtains ($j_\mu\equiv 2[Q,\partial_\mu Q]$)
in the lowest order in $\frac{1}{\lambda}$
\begin{equation}
\partial_\mu\psi_1+\epsilon_{\mu\nu}(j^\nu+2\gamma^\nu Q)=0,
\label{care1}
\end{equation}
whose current conservation is nothing but the equation of motion
(\ref{eqmt}).
The next lowest order produces
\begin{equation}
\partial_\mu\psi_2+(j_\mu+4\gamma_\mu Q)+
\epsilon_{\mu\nu}(j^\nu+2\gamma^\nu Q)\psi_1=0.
\label{care2}
\end{equation}
The higher order yields the following relations ($p\geq 0$);
\begin{eqnarray}
0&=&\partial_\mu\psi_{p+3}+\epsilon_{\mu\nu}
\sum_{m=0}^{[p/2]+1}\left(j^\nu+2(m+1)\gamma^\nu Q\right)
\psi_{p+2-2m}\nonumber\\
&+&\sum_{m=0}^{[\frac{p+1}{2}]}\left(j_\mu+4(m+1)\gamma_\mu Q\right)
\psi_{p+1-2m}+
2\epsilon_{\mu\nu}\gamma^\nu Q\sum_{m=0}^{[p/2]}(m+1)\psi_{p-2m},
\label{keep3}
\end{eqnarray}
which determines $\psi_n~(n\geq 3)$ completely in terms of  lower
$\psi_n$'s. 

The conserved charges can be constructed in two separate cases
depending on the boundary conditions.
One is the periodic boundary condition
$Q(x+2L)=Q(x)$. The conserved charges are given by
$M^{(n)}=\int_{-L}^{+L} J_0^{(n)}=\psi(-L)-\psi(+L)$.
(\ref{care1}) yields
\begin{equation}
M^{(1)}(Q)=\int_{-L}^{+L}dx (j^0(x)+2\gamma^0 Q(x)).
\label{keep1}
\end{equation}
The next order (\ref{care2}) produces
the nonlocal conserved charge as
\begin{eqnarray}
M^{(2)}(Q)&=&\int_{-L}^{+L}dx (j^1(x)+4\gamma^1 Q(x))
\label{keep2}
\\
&+&\int_{-L}^{+L}dx\int_{-L}^{+L}dx^\prime \theta(x-x^\prime)
(j^0(x)+2\gamma^0 Q(x))
(j^0(x^\prime)+2\gamma^0 Q(x^\prime)).\nonumber
\end{eqnarray}
The expressions for higher conserved charges
follow from (\ref{keep3}). 
In the rapidly decreasing case \cite{fadd}
with the boundary condition given by 
$\lim_{\vert x\vert\rightarrow \infty}
Q(x)=Q_0 (\neq 0)$, the above integrals
with $\pm L$ being replaced by $\pm\infty$
diverge in general. They become finite after
subtracting their values in the ground state, i.e.,
$M^{(n)}_{reg}=M^{(n)}(Q)-M^{(n)}(Q_0)$. 

In summary, starting from a discussion of soliton solutions of
simple $O(3)$ NLSM and an integrally extended version of it,
we verified that NLSM with the target space of  arbitrary HSS allows 
an integrable  extension by using the coadjoint orbit formulation.
The  covariant symplectic structure term on HSS results in the 
completely integrable equation of motion, 
and an explicit expression for infinite number of nonlocal currents
was obtained. A more detailed analysis on the subject of classical and 
quantum integrability such as Hamiltonian formulation and Poisson 
structure, current algebra, and multi-soliton solutions will be 
addressed elsewhere.

I thank  Q-H. Park for useful discussions.
This work is supported in part 
by the Korea Science and Engineering Foundation 
through the Center for Theoretical Physics  at 
Seoul National University, and the project number
(95-0702-04-01-3), 
and  by the Ministry of Education through the
Research Institute for Basic Science  (BSRI/97-1419).



\begin{thebibliography}{99}
\bibitem{com} Since our $\theta$ is defined to be $0\leq\theta\leq\pi$, 
$\xi$ must be restricted to be in the range $\xi_0\leq \xi<\infty$
in the solution (\ref{cru}).
\end{thebibliography}
\end{document}